\documentclass[a4paper,12pt]{article}
\usepackage[sumlimits,intlimits,namelimits]{amsmath}
\usepackage{amssymb}
\usepackage{a4wide}
\usepackage{comment}
\usepackage{mathtools}
\usepackage{cite}
\usepackage[linkcolor=blue,colorlinks=true]{hyperref}

\usepackage[english]{babel}
\usepackage{longtable, multirow,tabularx}
\makeatletter
\def\fmslash{\@ifnextchar[{\fmsl@sh}{\fmsl@sh[0mu]}}
\def\fmsl@sh[#1]#2{%
	\mathchoice
	{\@fmsl@sh\displaystyle{#1}{#2}}%
	{\@fmsl@sh\textstyle{#1}{#2}}%
	{\@fmsl@sh\scriptstyle{#1}{#2}}%
	{\@fmsl@sh\scriptscriptstyle{#1}{#2}}}
\def\@fmsl@sh#1#2#3{\m@th\ooalign{$\hfil#1\mkern#2/\hfil$\crcr$#1#3$}}
\makeatother
\usepackage{graphicx}
\usepackage{subfigure}
\usepackage{epstopdf}
\usepackage{xcolor}
 \usepackage{slashed}

\newcommand{\lbar}{\bar{\Lambda}}

\begin{document}
	\begin{titlepage}
		\begin{flushright}
            P3H-26-011 \\[0.2cm]
			SI-HEP-2026-01 \\[0.2cm]
            Nikhef-2026-002\\[0.2cm]
			\today
		\end{flushright}
		
		\vspace{1.2cm}
		\begin{center}
			\Large\bf
                Extending the Kinetic Mass \\[2mm] to Higher Orders in 
                \boldmath $1/m_Q$ \unboldmath 
		\end{center}
		
\vspace{0.5cm}
\begin{center}
{\sc Thomas Mannel} and {\sc Ilija S.\ Milutin}  \\[0.1cm]
{\sf Theoretische Physik 1,  Naturwiss. techn. Fakult\"at \\ 
Universit\"at Siegen  D-57068 Siegen, Germany} \\[0.5cm]
{\sc Rens Verkade} and {\sc K.~Keri Vos} \\[1mm]
{\sf Gravitational Waves and Fundamental Physics (GWFP), \\
Maastricht University, Duboisdomein 30, NL-6229 GT Maastricht, the Netherlands} \\ and \\
{\sf Nikhef, Science Park 105, NL-1098 XG Amsterdam, the Netherlands}
\end{center}
		
		\vspace{0.8cm}
		\begin{abstract}
			\vspace{0.2cm}\noindent 
            Currently, the kinetic mass is defined in terms of the pole mass and operators at order $1/m_Q^2$, which are known to N$^3$LO accuracy in $\alpha_s$. At the same time, the Heavy Quark Expansion (HQE) for inclusive semileptonic decays is known up to and including terms of order $1/m_Q^5$. Therefore, it is desirable to extend the definition of the kinetic mass to higher orders in $1/m_Q$. The original kinetic mass is based on the hadron-mass formula in Heavy Quark Effective Theory (HQET). However, the HQE is formulated in terms of matrix elements defined in full QCD to avoid the appearance of non-local matrix elements. To match this, we develop a definition of the kinetic mass rooted in full QCD. Starting from the hadron-mass formula derived from the energy-momentum tensor of full QCD, we define a relation between a general mass and the pole mass. Using a simple cut-off scheme, we compute a generalized kinetic mass at one loop to all powers of $1/m_Q$, which reproduces the well-known results for the kinetic mass up to $1/m_Q^2$. Our approach opens the road to a consistent use of the kinetic mass at higher-orders in the heavy quark expansion. 

		\end{abstract}

	\end{titlepage}

	\newpage
	\pagenumbering{arabic}

\section{Introduction}
Any perturbative calculation in QCD requires a proper scheme to fix the quark masses. In QED, with electrons and photons, all particles appear as asymptotic states, hence their pole masses can be defined through measurements. In QCD, the masses of quarks cannot be defined in this way due to quark confinement. 
At the same time, the pole mass is commonly used as a starting point for perturbative QCD calculations. 
It has been known for a long time that this leads to an ill-behaved perturbative series related 
to a renormalon problem in the pole mass. This renormalon excludes a precise 
extraction of a value of the pole mass from any observable, 
since it generates an intrinsic ambiguity in the pole mass of a few hundred MeV.

The solution to this problem is to switch to a ``short-distance mass'', which is not plagued by 
the renormalon problem and thus can be determined from data without an intrinsic ambiguity. The relation 
between a short-distance mass and the pole mass can be computed perturbatively, however, in terms of an 
ill-behaved series, reflecting the presence of the renormalon in the pole mass. 
 
A commonly used short-distance mass is the $\overline{\rm MS}$ scheme, which has become common 
for calculations at scales $\mu \gg m$ \cite{Chetyrkin:2010ic}. However, for heavy quarks with masses $m_Q$, there is still 
a region of scales $\mu \le m_Q$ which can be treated perturbatively, but where the $\overline{\rm MS}$ scheme
is not appropriate. 

To this end, other schemes have been identified, such as the potential-subtracted mass  
\cite{Beneke:1998rk} or the kinetic mass \cite{Bigi:1996si, Czarnecki:1997sz}, which can be used at scales 
$\mu \le m_Q$,  as well as schemes which can be used at all scales \cite{Hoang:2017suc}. Their definitions involve 
the removal of the infrared effects of the soft gluons with momenta smaller than a cut-off scale $\mu$ 
and are related to the colour Coulomb field of the 
heavy quark. This removal fixes the problems with infrared renormalons inherent in the definition of the pole mass.   
Furthermore, there is evidence that, in addition to the well-understood leading renormalon ambiguity 
of order $\Lambda_{\rm QCD}$, also sub-leading renormalons in the pole mass exist \cite{Beneke:1994sw}. These correspond to ambiguities of order $  \Lambda_{\rm QCD}  ( \Lambda_{\rm QCD} /  m_Q)^n$, $n = 1,2,...$, which become relevant when higher orders in $1/m_Q$ are studied.

In applications for inclusive heavy hadron decays, the Heavy Quark Expansion (HQE) has become a standard 
tool, which is based on an Operator Product Expansion (OPE), resulting in an expansion of decay rates 
in powers of $\Lambda_{\rm QCD} / m_Q$. This involves, on the one hand, non-perturbative matrix elements 
of local operators and, on the other hand, perturbative calculations of the Wilson coefficients of the OPE.
However, the perturbative expansion of the Wilson coefficients is also asymptotic; 
the corresponding power-like ambiguities relate to ambiguities in the definition of the 
non-perturbative matrix elements of the HQE \cite{Martinelli:1996pk}, such that the (differential) 
rates are free of ambiguities once the non-perturbative matrix elements and the quark mass are 
defined properly. 

In the present paper, we focus on mass definitions making use of a hard ``Wilsonian'' cut-off $\mu$.  
This can be implemented by cutting the absolute value of the spatial momentum $\vec{k}$ of the gluon \cite{Bigi:1996si,Bigi:1994ga}. 
In such a scheme, power-like divergencies become explicit, which are related to the renormalons appearing 
in the OPE. In dimensional regularization, which is the established regularization scheme, such power-like terms 
do not appear and thus it is not suitable for our purpose.    
Alternatively, one can also make use of a gradient-flow regularization, where the ``flow time'' acts as a hard cut-off, and which is manifestly gauge- and Lorentz
invariant \cite{Luscher:2011bx, Luscher:2013cpa}, and furthermore allows the calculation of higher orders in $\alpha_s$. We note that this has been applied recently to study the renormalon structure of perturbative 
series \cite{Beneke:2023wkq, Beneke:2025hlg}.

The relation of the kinetic mass to the pole mass has been calculated to order $\alpha_s^3$, which also contains the contributions of order $\mu^2/m_Q^2$ \cite{Fael:2020njb, Fael:2020iea}.  However, unlike the leading renormalon, these subleading pieces 
require including the non-perturbative matrix elements of the HQE at the same order 
to obtain a consistent treatment, i.e. one has to define the HQE parameters $\mu_\pi^2$ and 
$\mu_G^2$ accordingly. While this is the current state-of-the-art, the HQE for inclusive semileptonic 
processes is known already to $\Lambda_{\rm QCD}^3 / m_Q^3$ and even higher 
\cite{Dassinger:2006md,Mannel:2010wj,Mannel:2023yqf,Finauri:2025ost}. Therefore, the contributions 
to the kinetic mass of order $\alpha_s \mu^3/m_Q^3$ need to be included for a 
consistent evaluation of the inclusive rate.

The purpose of this paper is to discuss how to generalize the kinetic mass to extend it to higher orders in $1/m_Q$. The original kinetic mass is based on the relation between the hadron mass and the quark mass in HQET, which depends on non-local matrix elements starting at order $1/m_Q^3$. To avoid these, we work in full QCD and start from the hadron mass formula derived from the energy-momentum tensor in full QCD. This allows us to obtain a mass definition to all orders in $\mu/m_Q$. 

Our paper is organized as follows. We start by discussing the hadron to quark mass formula in HQET. In Sec.~\ref{sec:kinmass}, we set up the formulae for the mass definition in full QCD, 
thereby superseding the current definition of the kinetic mass. In Sec.~\ref{sec:pheno}, we discuss the application to the HQE of inclusive decays and some inconsistencies with the simple cut-off scheme and the small-velocity (SV) sum rules employed in the usual kinetic mass definition. We conclude with a summary and brief outlook. 

\section{Mass definitions and hadron mass formulae} \label{sec:massdef}

As pointed out in the introduction, the quark mass does not have any direct physical meaning, 
since its relation to observable quantities is non-perturbative. Thus, the quark mass is defined through 
a calculation scheme to compute an observable, from which a value for the mass in this specific scheme 
can be obtained, usually by a perturbative calculation. Consequently, different mass definitions can 
be related by a perturbative calculation. 

The pole mass plays a crucial role in heavy quark physics, since Heavy Quark
Effective Theory (HQET), as well as the Heavy Quark Expansion (HQE), are technically an expansion 
near the mass shell of the heavy quark, and so many calculations use the pole mass as a starting 
point, leading to ill-behaved perturbative series. 

The large order behaviour of this perturbative series is governed by infrared renormalons, which have been discussed for more than a decade, and have lead to various mass definitions which 
cure the renormalon problem of the on-shell mass \cite{Chetyrkin:2010ic,Beneke:1998rk, Bigi:1996si,Czarnecki:1997sz,Hoang:2017suc,Hoang:1998hm,Hoang:1998ng,Hoang:1999zc,Pineda:2001zq,Hoang:2008yj}.  All these mass schemes have in common that they require dealing with soft gluons in order to relate them to the pole mass. In the following sections, we shall discuss how this is typically done within the kinetic mass scheme.
However, before focusing on the kinetic mass, we start with some general remarks related to the on-shell quark mass $m_{os}$.

\subsection{Quark two-point function and on-shell quark mass} \label{sec:baremass}
The on-shell quark mass is usually related to the two-point vertex function of a quark:   
\begin{equation}
	\Gamma^{(2)} (p) = \fmslash{p} - m_0 - \Sigma(p)\ ,
\end{equation} 
where $m_0$ is the bare mass and $\Sigma(p)$ is the quark self-energy, which dresses the 
bare mass with gluons. The self-energy $\Sigma(p)$ is decomposed into its Dirac structure as 
\begin{equation} \label{SEDecomp}
	\Sigma (p) = \fmslash{p} \,  \Sigma_v (p^2) + m_0 \Sigma_s (p^2) \ ,
\end{equation} 
leading to
\begin{equation}\label{eq:two-point}
	\Gamma^{(2)} (p) = \fmslash{p} \left(1-\Sigma_v (p^2)\right) - m_0 \left(1+\Sigma_s (p^2)\right)\ ,
\end{equation}
The on-shell mass (or pole mass) is usually used as a starting point for any calculation, and it  
is defined as the location of the pole of the quark two-point function.  
Rewriting \eqref{eq:two-point} as    
\begin{equation}
	\Gamma^{(2)} (p) = [ \fmslash{p}-m_{os} ] \left(1-\Sigma_v (p^2)\right) 
	+ m_{os} \left(1-\Sigma_v (p^2)\right) -  m_0 \left(1+\Sigma_s (p^2)\right)\ ,
\end{equation}
and setting $\fmslash{p} = m_{os}$, gives the relation between 
the on-shell mass and the bare mass: 
\begin{equation} \label{OSdef}
	m_{os} = \frac{1+\Sigma_s (m_{os}^2)}{1-\Sigma_v (m_{os}^2)}\,  m_0  
	= m_0 \left(1+\Sigma_v (m_0^2) + \Sigma_s (m_0^2) + \cdots  \right)
	\equiv m_0 + \delta m_{os}\ ,
\end{equation}
such that, to order $\alpha_s$
\begin{equation} \label{DelMos}
	\delta m_{os} = m \left(\Sigma_v (m^2) + \Sigma_s (m^2) \right)\ .
\end{equation}
Note that since the self-energy terms are already of order $\alpha_s$, we do not specify which mass is to be used and we write a generic mass $m$.

In the following, we continue to work at ${\cal O}(\alpha_s)$ and consider only one-loop  diagrams. The self-energy, at one-loop order, then reads
 \begin{equation} \label{SelfE}
 	\Sigma(p) = - i C_F (4 \pi \alpha_s) \int \frac{\text{d}^4k}{(2 \pi)^4} \left(\frac{1}{k^2+i\epsilon}\right) 
 	\gamma_\mu \left(\frac{\fmslash{p}+\fmslash{k} + m}{(p+k)^2-m^2+i\epsilon}  \right) 
 	\gamma^\mu\ ,
 \end{equation}

which we decompose according to \eqref{SEDecomp}. We obtain 
\begin{align}
	 \Sigma_s (p^2,m^2) &= \frac{1}{4m} {\rm Tr} [ \Sigma(p) ] = - 4 i C_F (4 \pi \alpha_s) \int \frac{\text{d}^4k}{(2 \pi)^4} \left(\frac{1}{k^2+i\epsilon}\right) 
	\left(\frac{1}{(p+k)^2-m^2+ i \epsilon}  \right)\ ,   \nonumber \\  
	 \Sigma_v (p^2,m^2)     
	 &= \frac{1}{4p^2} {\rm Tr} [ \Sigma(p) \fmslash{p} ]  
	 \nonumber \\ 
	&=  2 i C_F (4 \pi \alpha_s) \int \frac{\text{d}^4k}{(2 \pi)^4} \left(\frac{1}{k^2+i\epsilon}\right)  
	\left(\frac{1}{(p+k)^2-m^2+ i \epsilon}  \right) \left[1+\frac{p\cdot k}{p^2} \right]\ . 
\end{align} 
These expressions can be reduced to master integrals, which are computed in the rest frame of the quark using a simple 
cut-off scheme where we restrict the spatial momentum of the gluon by $|\vec{k}| \le \Lambda$. 
We point out that this is a specific regularization scheme, which can be applied straightforwardly at one loop-order, 
and the results obtained for the matrix elements and for the mass renormalization depend on this scheme. However, 
choosing a different scheme with a hard cut-off will lead to different expressions, but the conclusions will remain 
the same. In particular, the observables will not depend on the chosen scheme.

Using \eqref{DelMos} and setting the external momenta on-shell, we find
\begin{eqnarray}\label{eq:dmos}
m_{os} - m_0 = \delta m_{os} = m C_F (4 \pi \alpha_s)  \left[ -2B_0(m^2,m^2)+ \frac{1}{m^2}\left(A_{01} - A_{02}(m^2) \right) \right],
\end{eqnarray} 
where $B_0, A_{01}$ and $A_{02}$ are given by: 

\begin{eqnarray}
	A_{01}  &=& i \int\frac{\text{d}^4k}{(2 \pi)^4} \left(\frac{1}{k^2+i\epsilon}\right) \ ,\label{A01}
	\\ 
	A_{02} (m^2)  &=& i \int\frac{\text{d}^4k}{(2 \pi)^4} 
	\left(\frac{1}{(p+k)^2 -m^2 + i \epsilon }  \right) \ ,\label{A02}
	\\ 
	B_0 (p^2,m^2) &=& i \int\frac{\text{d}^4k}{(2 \pi)^4} \left(\frac{1}{k^2+i\epsilon}\right) 
	\left(\frac{1}{(p+k)^2 -m^2 + i \epsilon }  \right) \ .\label{B0}
\end{eqnarray}
Finally, we obtain for the relation between the bare mass $m_0$ and the on-shell mass $m_{os}$
\begin{equation} \label{dmos}
	m_{os} - m_0 = \delta m_{os} = \frac{C_F \alpha_s}{2 \pi m} \left( \Lambda^2 - \Lambda \sqrt{\Lambda^2 + m^2}  +  
	3 m^2 \ln \left(\frac{\sqrt{\Lambda^2+m^2}+\Lambda}{m} \right) \right)\ ,  
\end{equation}
where $\Lambda$ is a generic cut-off.   
Note that \eqref{dmos} is valid for any value of the cut-off $\Lambda$. 
In case the self-energy were 
convergent, one would obtain a finite result as $\Lambda \to \infty$. The UV behaviour of \eqref{dmos} for $\Lambda \gg m$ 
is given by 
\begin{equation}
	\delta m_{os} =  m \left[ \frac{3 C_F \alpha_s}{4 \pi}\right]  \ln \left(\frac{\Lambda^2}{m^2} \right)  + \cdots\ ,  
\end{equation}
 where the ellipses denote constant terms or  terms suppressed by powers of $m/\Lambda$.
The expression in square brackets is just the anomalous dimension of the quark mass at 
one-loop, which appears here as the coefficient in front of the logarithmic divergence. 

We stress that in the above, the ``dressing'' 
of the bare mass $m_0$ involves all gluon momenta, in particular also the soft ones, 
which, however, are non-perturbative. Consequently, an obvious way to 
obtain a short-distance mass $\hat{m} (\mu)$ 
with only perturbative gluons is to cut away the soft gluons. This 
can be done straightforwardly by introducing
a cut-off $\mu$, $\Lambda_{\rm QCD} \ll \mu \ll m$, between the perturbative regime and the non-perturbative regime. Subtracting the infrared gluons with $|\vec{k}| \le \mu$ from the pole mass gives 
\begin{equation} \label{massRel}
m_{os} = m_0 + \delta m_{os} (\Lambda) =   \hat{m} (\mu) + \delta m_{os} (\mu) \ ,
\end{equation}
where the bare mass $m_0$ becomes dressed only with the perturbative gluons with momenta larger than $\mu$. Of course, the UV divergence is still present and has to be taken care of by mass renormalization. 
Using \eqref{dmos}, we can simply replace the generic cut-off with an IR cut-off  $\mu \ll m$ to obtain a formula for a short-distance mass containing all orders in $1/m_Q$:
\begin{eqnarray} \label{KinMassCut}
	m_{os} &=& \hat{m} (\mu) + \frac{C_F \alpha_s}{2 \pi m} \left( \mu^2 - \mu \sqrt{\mu^2 + m^2}  +  
	3 m^2 \ln \left(\frac{\sqrt{\mu^2+m^2}+\mu}{m} \right) \right) \ .
\end{eqnarray} 
Note that $\hat{m} (\mu)$ depends on the scheme we have chosen to eliminate the infrared gluons, but the 
dependence is eliminated by the second term in \eqref{KinMassCut}, such that the right hand side is independent 
of that scheme. We stress that the choice of a different way to cut off the infrared gluons will yield a different 
$\hat{m} (\mu)$, corresponding to a different second term in \eqref{KinMassCut}.

\subsection{From Hadron Mass to Quark Mass in HQET} 
The usual definition of the kinetic mass is derived from the relation of the 
heavy quark mass $m_Q$ to the hadron mass $M_H$ \cite{Bigi:1996si,Bigi:1994ga}. 
Using the Hamiltonian formulation of HQET, this reads 
\begin{align} 
M_H =\ m_Q \Bigg(&1 + \frac{\bar\Lambda}{m_Q} + \frac{\hat\mu_\pi^2  -  d_H  \hat\mu_G^2}{2 m_Q^2} 
+\frac{\hat\rho_D^3+d_H\hat\rho^3_{LS}}{2m_Q^3}
-\frac{\rho^3}{4m^3_Q}+ {\cal O} \left(\frac{\Lambda_{\rm QCD}^4}{m_Q^4} \right)  \Bigg) \ ,\label{HQETMassFomula}
\end{align}
where $m_Q$ is the mass of the heavy quark and the spin-counting factor $d_H$ is  $d_H = 1$  for the mesonic pseudoscalar ground 
states, $d_H = -1/3$ for vector 
meson ground state and $d_H=0$ for the $\Lambda$-type heavy baryon. The parameters $\bar\Lambda$, 
$\hat\mu_\pi^2$, $\hat\mu_G^2$, $\hat\rho_D^3$, $\hat\rho_{LS}^3$ and $\rho^3$ are non-perturbative matrix elements defined in terms of the static heavy-quark fields $h_v(x)$ and the infinite mass states $| \hat{H} (v) \rangle$ 
of HQET: 
\begin{align} 
\label{LamBar}
   \lbar &= \frac{  
   \langle 0 | \bar{q} \stackrel{\leftarrow}{(ivD)} \gamma_5 h_v | \hat{H}(v) \rangle}{\langle 0 | \bar{q}\gamma_5 h_v | \hat{H}(v) \rangle}  \ ,  \\
 \hat\mu_\pi^2\ &= -\frac{1}{2 M_H}  \langle \hat{H}(v) | \bar{h}_v (iD)^2 h_v | \hat{H}(v) \rangle     \  ,   \label{mupihat}\\ \label{muGhat}
 d_H \hat\mu_G^2 &= \frac{1}{2 M_H}  \langle \hat{H}(v) | \bar{h}_v (-i \sigma_{\mu \nu })(iD^\mu) (iD^\nu) h_v | \hat{H}(v) \rangle  \ ,\\ \label{rhoDhat}
 \hat{\rho}_D^3&= \frac{1}{4 M_H}\langle \hat{H}(v) | \bar{h}_v \big[iD_\mu,\big[iv \cdot D, iD^\mu \big] \big] h_v | \hat{H}(v) \rangle  \ , \\ 
 d_H\hat\rho_{LS}^3 &= \frac{1}{4 M_H}\langle \hat{H}(v) | \bar{h}_v \big\{ iD_\mu,\big[iv \cdot D, iD_\nu\big]\big\} (-i \sigma^{\mu\nu}) h_v | \hat{H}(v) \rangle  \ .  \label{rhoLShat}
\end{align}
and the parameter $\rho^3$ represents
a set of non-local matrix elements related to finite-mass 
corrections of the states, see e.g. \cite{Bigi:1994ga}. The quantities above are defined in HQET and are, therefore, independent of the heavy-quark mass. Furthermore, due to Heavy-Quark Spin Symmetry (HQS) $\hat\mu_G^2$ is identical for each member of a spin symmetry doublet, since the 
factor $d_H$ takes care of the proper spin counting.  
In \eqref{HQETMassFomula}, the left-hand side is a physical quantity, which does not depend on the definition of the quark mass. This means in turn that the
HQE parameters $(\bar\Lambda$, $\hat\mu_\pi^2,  \ldots)$ depend on the mass scheme such that the right-hand side becomes scheme independent as well. Since the pole mass of the quark is also
independent of the renormalization scale and gauge invariant, a mass definition can be obtained by replacing the left-hand side 
of \eqref{HQETMassFomula} by the pole mass.  Up to terms of order $1/m_Q^2$, the quark mass scheme on the right-hand side 
is then fixed by the computation of the parameters $\bar\Lambda$, 
$\hat\mu_\pi^2$ and $\hat\mu_G^2$ in that scheme. 
In order to make the kinetic mass a short-distance mass, a cut-off $\mu$ is introduced, which isolates
the infrared gluons in the (perturbative) calculations of the HQET matrix elements in \eqref{LamBar}-\eqref{muGhat}. 
In this way, we end up with 
the well-known expression defining the kinetic mass up to order $1/m_Q^2$.
\begin{equation}\label{eq:usualkinmass}
m_{os} = m_{\rm kin}(\mu) + [\bar\Lambda]_{\rm pert} (\mu) 
+ \frac{[\hat\mu_\pi^2]_{\rm pert} (\mu)}{2 m_Q} \ ,
\end{equation} 
where $[..]_{\rm pert} (\mu)$ denotes the calculation of the matrix elements in a 
perturbative scheme with infrared gluons below the scale $\mu$, such that $m_{\rm kin}$ involves only ultraviolet gluons. For an on-shell quark, we have $d_H=0$ and thus only the term induced by $\mu_\pi^2$ arises at $1/m_Q^2$. In the usual kinetic mass scheme, the perturbative quantities are then obtained using Small-Velocity (SV)
sum rules including an IR cut off.

We stress that this standard definition is based on the HQET mass formula. Starting at order $1/m_Q^3$, \eqref{HQETMassFomula} contains non-local matrix elements, combined together in the term $\rho^3$. Such non-local terms  
would also appear in the HQE, if one insists on fully expanding the mass dependence in powers 
of $1/m_Q$. However, the appearance of non-local matrix elements in the HQE can be circumvented by
formulating it in full QCD and not in the static limit. 
In this way, the HQE to any order in $1/m_Q$ can be derived in terms of local,  full QCD matrix elements. We note that all recent studies of the HQE for inclusive $b\to c$ semileptonic decays use this approach \cite{Fael:2024fkt, Bernlochner:2022ucr, Fael:2022frj, Finauri:2023kte, Finauri:2025ost, Carvunis:2025vab}. Therefore, it is desirable to find a relation between the hadron mass and the quark mass in terms of local full QCD matrix elements.

We can identify the full QCD counterparts of the HQET matrix elements by starting from a  multiplicative redefinition of the quark fields
\begin{equation}
Q_v (x) = \exp(i m_Q (vx)) Q(x) \ ,
\end{equation}
with $Q(x)$ being the heavy quark field in full QCD and use the 
states $|H(v) \rangle$ of full QCD to define  
\begin{eqnarray}
&&  \mu_\pi^2\ = -\frac{1}{2 M_H}  \langle H(v) | \bar{Q}_v (iD)^2 Q_v | H(v) \rangle\ , \qquad \hat\mu_\pi^2 = \lim_{m_Q \to \infty} \mu_\pi^2   \  , \label{eq:QCDmupi}   \\  \label{eq:QCDmuG}
&&   \mu_G^2 = \frac{1}{2 M_H}  \langle H(v) | \bar{Q}_v (-i \sigma_{\mu \nu })(iD^\mu) (iD^\nu) Q_v | H(v) \rangle \ , \qquad d_H \hat{\mu}_G^2 = \lim_{m_Q \to \infty} \mu_G^2  \ ,
\\
&& |\hat{H}\rangle =\lim_{m_Q\rightarrow\infty} |H\rangle \ .
\end{eqnarray}
Note that to leading order in $1/m_Q$, these matrix elements coincide with the HQET matrix 
elements. However, the matrix elements $\mu_\pi^2$ and $\mu_G^2$ depend 
in a non-trivial way on $m_Q$, 
making them different from their HQET versions at subleading order in $1/m_Q$. 
Note also that for a heavy Lambda baryon ($\Lambda_Q$), where $d_H=0$,  we have  
\begin{eqnarray}
\frac{1}{2 M_{\Lambda_Q}} \langle \Lambda_Q (v) | \bar{Q}_v (-i \sigma_{\mu \nu })(iD^\mu) (iD^\nu) Q_v | \Lambda_Q (v) \rangle  = \mu_G^2 (\Lambda_Q) \sim  \frac{\Lambda_{\rm QCD}^3}{m_Q} 
\end{eqnarray}
The first step to extend the kinetic mass
definition to full QCD is to switch to a hadron-mass formula in full QCD. In the next section, we will show how this can be achieved.

\section{Extended kinetic mass from the QCD mass formula}
\label{sec:kinmass}

The starting point is the   
(symmetric) energy momentum tensor $\Theta_{\mu \nu}$ of QCD 
which satisfies 
\begin{equation}
	\partial^\mu  \Theta_{\mu \nu} = 0 \ ,
\end{equation}
due to energy-momentum conservation. From this, we can derive the normalization condition 
\begin{equation} \label{ThetaNorm}
	\langle H(p) |  \Theta_{\mu \nu}  | H(p) \rangle = \langle    \Theta_{\mu \nu}  \rangle = 2 p_\mu p_\nu\ ,
\end{equation}
which holds for any  
hadronic state $ | H(p) \rangle$ defined in full QCD. 

Taking the trace of \eqref{ThetaNorm} yields a formula relating the forward matrix element of  $\Theta_\mu^{\,\,\,\mu }$ to the 
mass of the hadron $H$  :
\begin{equation} \label{forward}
\frac{1}{2 M_H}	\langle H(p) |  \Theta_\mu^{\,\,\,\mu }  | H(p) \rangle \equiv \frac{1}{2 M_H} \langle 	   \Theta_\mu^{\,\,\,\mu }  \rangle = M_H \ ,
\end{equation}
where we introduced the shorthand notation $\langle\dots\rangle\equiv\langle H(p)|\dots|H(p)\rangle$.

On the other hand, we can compute $\Theta_\mu^{\,\,\,\mu }$ in terms of quark and gluon fields. Starting from the 
QCD Lagrangian, 
\begin{equation}
{\cal L} = \bar{Q} (i \fmslash{D} - m_Q (\mu)) Q + \sum_q	\bar{q} (i \fmslash{D} - m_q (\mu)) q - \frac{1}{4} 
G_{\mu \nu}^a G^{\mu \nu , a}  + \ldots\ ,
\end{equation} 
where we have singled out the heavy quark $Q$, $iD_\mu = i \partial_\mu + g_s (\mu) A_\mu^a T^a$  and $ig_s(\mu)G_{\mu\nu}=[D_\mu, D_\nu]$ is the gluon field strength tensor with $G_{\mu\nu} = G_{\mu\nu}^a T^a$. The trace of the 
energy momentum tensor becomes
\begin{eqnarray}
\Theta_\mu^{\,\,\,\mu } &=& - m_Q \frac{\partial}{\partial m_Q} {\cal L} + \left(\mu \frac{\partial}{\partial \mu} - 
 m_q \frac{\partial}{\partial m_q}   \right) {\cal L}  \\ 
 &=& \left[m_Q (\mu) - \mu \frac{d m_Q (\mu)}{d \mu} \right] \bar{Q} Q +  
 \frac{\beta(\alpha_s)}{4 \alpha_s}  G_{\mu \nu}^a G^{\mu \nu, \, a} 
 + \sum_q m_q (1+\gamma_{m_q}) \bar{q}q \ ,
\end{eqnarray}  
where $\gamma_{m_q}$ is the anomalous dimension of the light quark mass
\begin{equation}
m_q \gamma_{m_q} = \mu \frac{\text{d} m_q (\mu)}{\text{d} \mu}\ ,
\end{equation}
and $\beta(\alpha_s)$ is  
\begin{align}
    \beta(\alpha_s)=\mu\frac{\text{d} \alpha_s(\mu)}{\text{d}\mu}=\mathcal{O}(\alpha_s^2)\ . \label{beta_eq}
\end{align}

Setting the light quark masses to zero,  
we find (the factor $2M_H$ is the normalization factor of the hadronic states)
\begin{equation}  \label{QCDMassFormula}
 \frac{ \langle    \Theta_\mu^{\,\,\, \mu}  \rangle }{2 M_H} = M_H  =   m_Q(\mu) \frac{\langle \bar{Q} Q \rangle}{2 M_H} - \mu \frac{\text{d} m_Q(\mu)}{\text{d} \mu}   \frac{\langle \bar{Q} Q \rangle}{2 M_H} 
	+  \frac{\beta(\alpha_s(\mu))}{4 \alpha_s(\mu)}  \frac{\langle G_{\mu \nu}^a G^{\mu \nu, \, a} \rangle}{2 M_H} \ .
\end{equation} 

This is a well-known formula for the mass of heavy hadrons (see eg.~\cite{Novikov:1981xi}), 
where all matrix elements are defined in full QCD. In particular, 
the left-hand side is an observable, while on the right-hand side we have QCD entities, 
which depend on the scheme
as well as on the renormalization scale $\mu$, so all these dependencies have to cancel. 
Furthermore,  this relation is exact to all orders in $1/m_Q$, where, however, the matrix 
elements on the right-hand side depend in a complicated way on $m_Q$.

In the same spirit as for the original kinetic mass definition, we now use \eqref{QCDMassFormula}
to explicitly define an extended kinetic mass $m_{\rm kin} (\mu)$ by 
\begin{eqnarray} \label{extkin}
m_{os} &=& \left(m_{\rm kin} (\mu) - \mu \frac{\text{d} m_{\rm kin} (\mu)}{\text{d} \mu} \right) \left[\frac{\langle  \bar{Q} Q \rangle}{2 M_H} \right]_{\rm pert} \!\!\!\!\!(\mu)+  \frac{\beta(\alpha_s(\mu))}{4 \alpha_s(\mu)}  \left[ \frac{\langle G_{\mu \nu}^a G^{\mu \nu, \, a} \rangle}{2 M_H} \right]_{\rm pert} \!\!\!\!\! (\mu) \ ,
\end{eqnarray}
where $[...]_{\rm pert} (\mu)$ denotes a perturbative calculation of the corresponding matrix element in a hard cut-off scheme, this time in full QCD. Similarly to the discussion following \eqref{KinMassCut}, the kinetic-mass scheme is defined here by the precise definition 
of the hard cut-off scheme used to calculate the matrix elements indicated by $[\cdots]_{\rm pert}$. 
In the following, we illustrate this at order $\alpha_s$ in the simple cut-off scheme introduced above,  
by cutting the absolute value of the three-momenta of the gluons, as originally proposed 
in \cite{Bigi:1996si}. We will show that this simple scheme is sufficient to compute the kinetic mass to all orders in $1/m_Q$ at one loop. Furthermore, this reproduces the results obtained in the HQET calculation up to order $1/m_Q^2$. However, we emphasize that 
\eqref{extkin} can be used for any other cut-off scheme.

\subsection{Calculation of the on-shell $\bar{Q}Q$ matrix element}\label{sec:QQ}
First, we further specify our mass scheme, introduced in general in \eqref{extkin}, by defining how to evaluate the perturbative matrix elements. As in the HQET kinetic mass definition, we introduce a cut-off $\mu$ on the absolute value of the gluon three-momentum $\vec{k}$. However, we now evaluate the matrix elements with on-shell free quarks in full QCD. To indicate this scheme, we adopt the following notation : 
\begin{equation}
[\langle ... \rangle]_{\rm pert} (\mu) \equiv	\langle Q(p,s) | \cdots | Q(p,s) \rangle_\mu\ = \langle ... \rangle_\mu , \quad |\vec{k}| \leq \mu \ , \label{eq:def_cutoff}
\end{equation} 
where $Q$ is a free, on-shell quark. Furthermore the states are now normalised using the on-shell mass.

We proceed by calculating the diagonal matrix element  
\begin{equation}
M_{\bar{Q}Q} \equiv     \langle  \bar{Q} Q   \rangle _{\mu} \end{equation}
with on shell quarks, $\fmslash{p} = m_{os}$. 
At tree-level, this matrix element is just  
\begin{equation}
M_{\bar{Q}Q}^{(0)} = \bar{u}(p,s) u(p,s) = 2 m_{os}\ .
\end{equation}
\begin{figure}
\centering     
\subfigure[]{\label{fig:QQ_a}\includegraphics[width=0.3\textwidth]{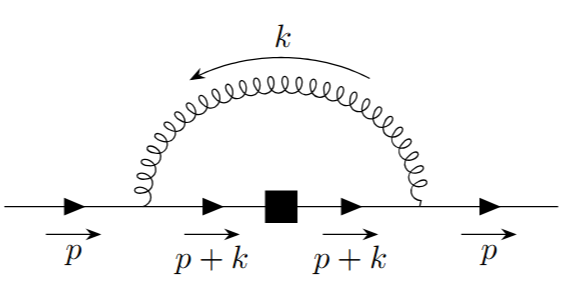}}
\subfigure[]{\label{fig:QQ_b}\includegraphics[width=0.29\textwidth]{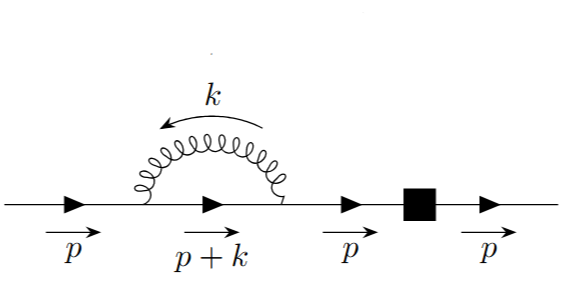}}
\subfigure[]{\label{fig:QQ_c}\includegraphics[width=0.305\textwidth]{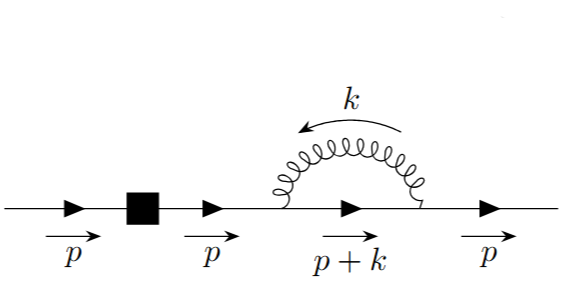}}
\caption{The one-loop diagrams contributing to $M_{\bar{Q}Q}$, where (b) and (c) correspond to the external field renormalization. The black square represents the insertion of $\bar{Q}Q$. }
\label{fig:QQinsertion}
\end{figure}
At one loop, we need to compute the Feynman diagrams shown in Fig.~\ref{fig:QQinsertion}. Specifically, Fig.~\ref{fig:QQ_a} corresponds to
\begin{equation}
	 - i C_F (4 \pi \alpha_s) \int \frac{\text{d}^4k}{(2 \pi)^4} \frac{1}{k^2+i\epsilon}\,\bar{u}(p,s)\left( 
	\gamma_\mu\, \frac{\fmslash{p}+\fmslash{k} + m}{(p+k)^2-m^2+i\epsilon}\,  \frac{\fmslash{p}+\fmslash{k} + m}{(p+k)^2-m^2+i\epsilon}\, \,   
	 \gamma^\mu\right) u(p,s)\ .\label{gamma}
\end{equation} 
We note that \eqref{gamma} can be written as 
\begin{equation}
	\bar u(p,s)\left(\frac{\partial}{\partial m} \Sigma(p) \right)u(p,s)\ ,
\end{equation}
where $\Sigma (p)$ is the self-energy at one loop given in \eqref{SelfE}.

To obtain the full $\bar{Q}Q$ matrix element, we include the self-energy insertions 
on the external legs given in Figs.~\ref{fig:QQ_b} and \ref{fig:QQ_c}, which renormalise the mass and wave function:  
\begin{eqnarray} 
	M_{\bar{Q}Q}^{(1)} &=& \bar{u}(p,s) \left(1+ \left[\frac{\partial}{\partial m} +    \frac{\partial}{\partial \fmslash{p}} \right] \Sigma(p) \right) u(p,s) 
	\nonumber\\ 
	&=& 
	\bar{u}(p,s) \left( 1+\left[\frac{\partial}{\partial m} + 
	\frac{p_\mu}{m}
	   \frac{\partial}{\partial p_\mu} \right] \Sigma(p) \right) u(p,s)\ , \label{OSM}
\end{eqnarray}
where $M^{(1)}_{QQ}$ is the result to one loop, including the tree-level contribution. Using the derivation of the self-energy, we obtain   
\begin{eqnarray}\label{Eq:QQ_master}
	\frac{1}{2m}\langle \bar{Q} Q \rangle &=& 1- C_F (4 \pi \alpha_s) \left( 
	2 m \left[ \left. \frac{\partial}{\partial m} B_0 (p^2,m^2) \right|_{p^2 = m^2}
	+    
	\left. \frac{p_\mu}{m}\frac{\partial}{\partial p_\mu} B_0 (p^2,m^2) \right|_{p^2 = m^2} 
	\right] \right. \nonumber \\  
	&&  \quad \quad
	\left. + 2 B_0 (m^2,m^2) + \frac{1}{m^2} A_{01} + \frac{1}{m} \frac{\partial}{\partial m} A_{02} (m^2) - \frac{1}{m^2}A_{02} (m^2)  
	\vphantom{\left. \frac{\partial}{\partial m} B_0 (p^2,m^2) \right|_{p^2 = m^2} }\right)\  ,
\end{eqnarray}
in terms of the master integrals given in equations \eqref{A01}-\eqref{B0}.
Finally, by computing the master integrals with a cut-off $\mu$ on the gluon 3-momentum, we obtain at one loop  
\begin{align} 
\frac{1}{2m}\langle \bar{Q}& Q \rangle_\mu = 1   \nonumber \\
&- C_F \frac{\alpha_s}{2 \pi} \left[ 
\frac{\mu^2}{m^2} + 3 \frac{\mu}{\sqrt{\mu^2+m^2}} - \frac{\mu^3}{m^2 \sqrt{\mu^2+m^2}}
- 3 \ln\left(\frac{\sqrt{\mu^2+m^2}+\mu}{m} \right) 
 \right] \ . \label{QQco}
\end{align}
For convenience, we define 
\begin{equation}\label{eq:QQmu}
\frac{\langle  \bar{Q}Q \rangle_{\mu} }{2 m} 
\equiv 1 + \Delta_{\bar{Q}Q} (\mu)\ , 
\end{equation}
where $\Delta_{\bar{Q}Q} (\mu)$ is the one-loop expression. Expanding around $\mu \ll m$, gives
\begin{equation} \label{QQExp}
\frac{1}{2m}\langle \bar{Q} Q \rangle _\mu 
= 1-C_F \frac{\alpha_s}{2 \pi} \frac{\mu^2}{m^2} + C_F \frac{\alpha_s}{\pi} \frac{\mu^3}{m^3} 
 - \frac{7}{10} C_F \frac{\alpha_s}{\pi} \frac{\mu^5}{m^5} + \cdots\ ,
\end{equation}
while the case $\mu \gg m$ yields 
\begin{equation}
	\frac{1}{2m}\langle \bar{Q} Q \rangle_\mu = 1 
	+  \left[\frac{3 C_F \alpha_s}{4 \pi} \right] \left(\ln\left(\frac{\mu^2}{m^2} \right) + \cdots \right) \ ,
\end{equation}
where the ellipses denote constants and terms suppressed by powers of $1/m$. We note that the logarithmic term is again related to the ultraviolet behaviour of the 
$\langle \bar{Q}Q \rangle$ matrix element. Its prefactor is the one-loop anomalous dimension of the 
mass, which is natural, since the combination $m(\mu) \langle \bar{Q}Q \rangle_\mu $ 
has to be UV convergent.

In fact, we note that the diagonal matrix element of $\bar{Q}Q =\bar{Q}_v Q_v  $ with a heavy-hadron state $| H(v) \rangle $ 
moving with the four-velocity $v$ reads to all orders in $1/m_Q$ (see e.g. \cite{Mannel:2018mqv})
\begin{equation}\label{eq:QQ_full}
\frac{1}{2 M_H} \langle H(v) | \bar{Q} Q | H(v) \rangle =  1 - \frac{1}{2m_Q^2} (\mu_\pi^2 - \mu_G^2) \ ,
\end{equation}
where the HQE parameters in full QCD are defined in \eqref{eq:QCDmupi} and \eqref{eq:QCDmuG}. Introducing a cut-off $\mu$ and calculating the matrix element perturbatively, we obtain
\begin{equation} 
\frac{1}{2m}\langle \bar{Q} Q \rangle _\mu = 1 - \left(\frac{[\mu_\pi^2]_{\rm pert}}{2 m^2} - \frac{ [\mu_G^2]_{\rm pert}}{2 m^2}\right)\ ,\label{QQasmupimuG}
\end{equation}
to all orders in $1/m$. As a check of our result in \eqref{QQco}, we can compute $\mu_\pi^2$ and $\mu_G^2$ (defined in \eqref{eq:QCDmupi} and \eqref{eq:QCDmuG}) in full QCD with a Wilsonian cut-off. Using the QCD Feynman rules, we find 
\begin{align}
    [\mu_\pi^2]_{\rm pert}(\mu)&=C_F\frac{\alpha_s}{\pi}\left[\mu^2+\frac{6\mu m^2+2\mu^3}{\sqrt{\mu^2+m^2}}+6m^2\ln\left(\frac{\sqrt{\mu^2+m^2}-\mu}{m}\right)\right]\ , \label{eq:mupi_pert}\\
    [\mu_G^2]_{\rm pert}(\mu)&=  C_F\frac{\alpha_s}{\pi}\left[\frac{3\mu m^2+3\mu^3}{\sqrt{\mu^2+m^2}}+3m^2\ln\left(\frac{\sqrt{\mu^2+m^2}-\mu}{m}\right)\right]\ ,\label{eq:muG_pert}
\end{align}
at order $\alpha_s$ but to all orders in $1/m$. Inserting the above results into \eqref{QQasmupimuG} reproduces the direct cut-off calculation of $\langle \bar{Q}Q\rangle$ in \eqref{QQco}.

Using a Taylor expansion in $1/m$, we obtain 
\begin{align}
[\mu_\pi^2 ]_{\rm pert}(\mu) &=   \frac{C_F\alpha_s}{\pi} \left(\mu^2 +\frac{4\mu^5}{5 m^3} +\mathcal{O}(\mu^7)\right)\ ,\\ 
[\mu_G^2 ]_{\rm pert} (\mu)&=  \frac{C_F\alpha_s}{\pi}\left(\frac{2\mu^3}{m}-\frac{3 \mu^5}{5 m^3}+ \mathcal{O}(\mu^7)\right) \ .\ 
\end{align}
We note that for the free quark $d_H=0$, therefore, contributions to $[\mu_G^2]_{\rm pert}$ only arise at subleading order in $\mu/m$. 

\subsection{Calculation of the anomaly terms}
Having dealt with the first term in \eqref{extkin}, it is now time to turn to the two remaining terms:
\begin{equation}
- \left( \mu \frac{\text{d}}{\text{d} \mu} m_{\rm kin}(\mu) \right) \frac{\langle  \bar{Q}Q  \rangle_{\mu} }{2 m_{os}}
  + \frac{\beta(\alpha_s(\mu))}{4 \alpha_s (\mu)}  \frac{1}{2 m_{os}}\left\langle   
G_{\mu \nu}^a G^{\mu \nu, \, a}  \right\rangle_{\mu} \ ,  
\end{equation}
where the matrix elements are calculated following our scheme definition in \eqref{eq:def_cutoff}. 
In the classical formula for the trace of the energy-momentum tensor, these terms do not appear. The quantum version picks up the additional terms which are present due to the dependence on the cut-off or (after renormalization) 
the renormalization scale $\mu$.

Let us start with the anomalous dimension of the heavy quark mass. Since the tree-level  quark mass is scale independent, the leading contribution of this term is of order $\alpha_s$. As a result, only the  tree-level contribution to $\bar{Q}Q$ contributes to the overall order $\alpha_s$ result. Using the result of \eqref{QQco}, the first term can be readily included. 

\begin{figure}
    \centering
    \includegraphics[width=0.5\linewidth]{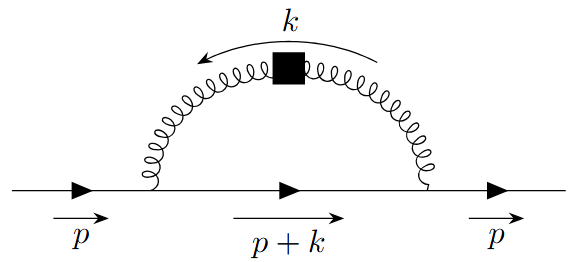}
    \caption{The Feynman diagram for the on-shell matrix element from the $G_{\mu\nu}^a G^{\mu\nu,a}$ operator, denoted by the black square.}
    \label{Gsquared}
\end{figure}

The second term emerges from the $\mu$-dependence of $\alpha_s$. The contribution of the $\langle GG \rangle$ operator to the on-shell matrix element of a quark is depicted in Fig.~\ref{Gsquared}, which starts at order $\alpha_s$. In addition, the prefactor of the 
matrix element, $\beta(\alpha_s) / \alpha_s$, also starts at $\alpha_s$ as can be seen from \eqref{beta_eq}. This pushes the total contribution of the second term to order $\alpha_s^2$. Since we compute only to order $\alpha_s$, this term does not contribute. The extended kinetic mass definition \eqref{extkin}, at one loop, then simplifies to

\begin{equation}\label{eq:mos}
    m_{os} = \left(1+\Delta_{\bar{Q}Q}\right)m_{\rm kin} (\mu) - \mu \frac{\text{d} m_{\rm kin} (\mu)}{\text{d}  \mu}  \ ,
\end{equation}
where $\Delta_{\bar{Q}Q}$ is defined in \eqref{eq:QQmu} and can be read off from \eqref{QQco}.

\subsection{The extended kinetic mass}\label{subsec:equiv}

The extended kinetic mass can be obtained by solving the differential equation \eqref{eq:mos}. Taking the general ansatz of the relation between the pole mass and a short-distance mass in \eqref{massRel},
\begin{equation}
m_{\rm kin} = \hat{m} (\mu) \equiv  m_{os} - \delta m_{os} (\mu)\ ,
\end{equation}
and inserting this into \eqref{eq:mos}, gives
\begin{equation} \label{DiffEq}
m_{os} \Delta_{\bar{Q}Q} (\mu) - \delta m_{os} (\mu) + \mu \frac{\partial }{\partial \mu} \delta m_{os} (\mu) = 0\ .
\end{equation}
Using the result for $\Delta_{\bar{Q}Q}$ from \eqref{QQco} and solving the differential equation, we find
\begin{equation} \label{diffSol}
\delta m_{os} = \frac{C_F \alpha_s}{2 \pi m} \left( \mu^2 - \mu \sqrt{\mu^2 + m^2}  +  
3 m^2 \ln \left(\frac{\sqrt{\mu^2+m^2}+\mu}{m} \right) \right) + a \frac{C_F \alpha_s}{\pi} \mu \ .
\end{equation}
Here $a$ is an arbitrary integration constant, which leaves an ambiguity in the order $\mu$ terms at all orders in $\alpha_s$.  

Comparing this result with the expression in \eqref{KinMassCut} obtained from the calculation of the self-energy diagram, we note that they match for $a = 0$. 
Therefore, the extended kinetic mass at one loop is equivalent to the short-distance mass obtained by cutting out the soft gluon momenta from the pole mass, using the simple 
cut-off scheme defined above. However, the leading power in $\mu$ remains ambiguous when 
choosing the definition via the mass formula \eqref{extkin}, while 
the calculation using the cut-off self-energy diagram singles out the value $a=0$. Expanding our closed formula for the extended kinetic mass in powers of $\mu \ll m$, gives
\begin{equation}\label{eq:mkinQCD}
    m_{\rm kin} (\mu) = m_{\rm os} -  
 \frac{C_F \alpha_s}{\pi} \mu - \frac{C_F \alpha_s}{\pi}  \frac{\mu^2}{2m}  + \frac{C_F \alpha_s}{\pi} \frac{\mu^3}{2m^2}+ \mathcal{O}\left(\frac{1}{m^3}\right)\  .
\end{equation}

Comparing with the usual HQET kinetic mass, implemented in HQE applications through \cite{Bigi:1996si,Bigi:1994ga, Fael:2020njb} 
\begin{equation}
m_{\rm kin}^{\rm HQET} (\mu) = m_{\rm os} - \frac{4}{3} \frac{C_F \alpha_s}{\pi} \mu -\frac{C_F \alpha_s}{\pi} \,  \frac{\mu^2}{2m} \ ,
\end{equation}
we note that the first two terms agree, aside from the ambiguous leading power in $\mu$. In the HQET kinetic mass this ambiguous term is obtained from the first moment of the spectral function of semileptonic $b \to c$ transitions. We note that in the shape-function scheme \cite{Bosch:2004th}, we have 
$
 m_{\rm SF} (\mu) = m_{os} -   (C_F \alpha_s /\pi)\, \mu  + \ldots \ , 
$ 
which agrees with our definition in \eqref{eq:mkinQCD} at leading power.

The ambiguity in the leading term can be traced back to the fact that the absence 
of $1/m$ contributions in the HQE is only true if one uses the pole mass as the expansion 
parameter. Otherwise, a residual mass term $\delta m = m_{os} - m$ appears. In \cite{Falk:1992fm}, it has been 
shown that in the description of physical matrix elements, only the combination 
$\bar\Lambda - \delta m$ appears, 
this combination 
is defined in \eqref{LamBar} for the choice of $\delta m =0 $. This combination is invariant under a redefinition of the linear term of the form 
$\delta m \to \delta m + \kappa \mu$ and simultaneously 
$\bar\Lambda \to \bar\Lambda + \kappa \mu$.
This means that no matrix element in the OPE can be used to pin down the leading term in $\delta m$ because it is fully degenerate with $\bar \Lambda$. The only combination that appears ($\bar\Lambda - \delta m$) is independent of the choice of $m_Q$ and can not be used to determine it.
This reasoning has been applied to the leading renormalon in \cite{Neubert:1994wq}, with the overall 
conclusion that the linear term in $\mu$ is arbitrary, and this ambiguity is resolved 
as soon as the mass is determined from an observable (see e.g. \cite{Penin:1998kx,Penin:1998wj, Boushmelev:2023kmf}).

\section{Kinetic mass and the HQE}\label{sec:pheno}
One of the main applications of the kinetic mass is the HQE for inclusive processes. As inputs for the HQE we need - aside from the mass, 
which appears to the fifth power in the expressions for the total rate - also the HQE parameters. These are defined 
in terms of forward matrix elements of local operators, some of which also appear in the mass definitions 
discussed above. 

However, the large-order behaviour of the perturbative 
series of the Wilson coefficients of the HQE depends on the scheme used to define the quark mass and the HQE parameters. It is well known that perturbative series of the Wilson coefficients are asymptotic, since they
exhibit a factorial growth at large orders, which can be related to singularities in their Borel transform 
known as infrared renormalons. These correspond to ambiguities in powers suppressed contributions, i.e. in the 
mass and in the HQE parameters, which can be resolved by fixing the mass and the HQE parameters using data 
\cite{Penin:1998kx, Penin:1998wj,Boushmelev:2023kmf}.

The leading renormalon, which is related to the ambiguity of the term 
of order $\mu$, is entirely taken care of by a proper definition of the quark mass, since the HQE has no term at 
order $1/m_Q$. However, extending the kinetic mass to terms of order $\mu^2/m$ requires a corresponding definition 
for the HQE parameters $\mu_\pi^2$ and $\mu_G^2$ to take into account the interplay between the subleading renormalon 
in the mass and the renormalons in the HQE parameters to obtain a consistent treatment. 

To this end, changing from the pole-mass scheme to the extended kinetic scheme requires redefining the HQE parameters. We point out that 
the usual redefinition: 
\begin{eqnarray}
\left[ \mu_\pi^2\right]_{\rm kin} (\mu) &=&  \left[ \mu_\pi^2\right]_{\rm pole} + \left[\mu_\pi^2 \right]_{\rm pert} (\mu)\ , \label{mupikin} 
\end{eqnarray}
where $[...]_{\rm pert}$ denotes the perturbative calculation of the corresponding operator in the cut-off scheme, takes care only of the subleading renormalon in the pole mass. 

In fact, looking at the mass definition in HQET \eqref{HQETMassFomula}, 
the mass parameter on the right-hand side depends on the mass definition, 
and this dependence is cancelled by the corresponding definition of the HQE parameters. 
Writing down \eqref{HQETMassFomula} for the pole mass and the kinetic mass 
up to order $1/m^2$ one derives \eqref{mupikin} by inserting the relation between the pole mass and the kinetic mass . However, the dependence on $\mu_\pi^2$ of the (differential) rates is not only due to the fact that 
the observables should depend only on hadronic quantities, 
in particular the true phase space boundaries will depend on the 
hadron masses. From this we conclude that using \eqref{mupikin} only partially removes the first subleading renormalon of the HQE.

We note that in most of the current applications, also $\rho_D^3$ and $\rho_{LS}^3$ are redefined analogously  by subtracting their perturbative part, but
in the light of the above discussion this seems to be inconsistent, as long as $m_{\rm kin}$ is not extended to 
order $\mu^3/m^2$. 

However, the simple cut-off scheme used above is insufficient to calculate the perturbative corrections to the HQE parameters at $1/m_Q^3$, 
i.e. to $\rho_D^3$ and $\rho_{LS}^3$, and higher. Starting at this order, the integrals computed in full QCD exhibit 
power-like UV divergences, complicating the separation of the IR  
and UV contributions. Developing a procedure to properly disentangle IR and UV contribution 
lies beyond the scope of this paper. We note, however, that the gradient flow method could be a promising 
method to provide a cut-off scheme 
which allows this separation. A first application along these lines can be found in \cite{Beneke:2023wkq,Beneke:2025hlg}. 

The simple cut-off scheme works up to $1/m_Q^2$ since at this order only the operators of lowest dimension appear. The typical problem with a hard cut-off scheme is that at higher orders mixing with lower-dimensional operators  occurs, reflecting the fact UV and IR contributions are challenging to disentangle. 

However, we may still define the 
perturbative contributions to $\rho_D^3$ and $\rho_{LS}^3$ in the simple cut-off scheme used above by exploiting the relations obtained from the equations
of motion. Specifically, the HQE operators in (\ref{eq:QCDmupi}, \ref{eq:QCDmuG}) defined with the full 
covariant derivative are also often defined in terms of $iD^\perp$, which contains only the spatial parts of the 
covariant derivative and is given by
\begin{equation}
    iD_\mu^\perp = iD_\mu - v_\mu(ivD) \ .
\end{equation}
Relations between the operators defined with $iD_\mu$ and those defined with $iD^\perp_\mu$ based on the equations of motions are given in \cite{Mannel:2018mqv, Mannel:2023yqf}. This gives
\begin{equation}\label{eq:perp_rel}
(\mu_G^2)^\perp = \mu_G^2 + \frac{\rho_D^3+ \rho_{LS}^3}{m}  \ ,
\end{equation}
to all orders in $1/m$ and $\alpha_s$. Here, $(\mu_G^2)^\perp$ is defined analogous to $\mu_G^2$, however, including only the spatial parts of the covariant derivative. Since $(\mu_G^2)^\perp$ enters at $1/m_Q^2$, we can calculate its perturbative part in the simple cut-off approach in full QCD. We obtain
\begin{align}
\left[(\mu_G^2)^\perp\right]_{\rm pert}&=\frac{C_F\alpha_s}{\pi}\left[\mu\sqrt{\mu^2+m^2}+\frac{m^2}{2}\log\left(\frac{\sqrt{\mu^2+m^2}-\mu}{\sqrt{\mu^2+m^2}+\mu}\right)\right]\nonumber\\
    &=\frac{C_F\alpha_s}{\pi}\frac{2\mu^3}{3m}+\mathcal{O}(\mu^4)\ ,
\end{align}
which does not have a leading term in $\mu^2$. Similarly,   $[\rho_{LS}^3]_{\rm pert}$ also has no leading term in $\mu^3$ and starts only at $\mathcal{O}(\mu^4/m)$. 

Therefore, we can use the perturbative 
calculations of $(\mu_G^2)^\perp$ and $\mu_G^2$ and \eqref{eq:perp_rel} to determine the perturbative correction to $\rho_D^3$. This leads to
\begin{equation}
\left[ \rho_D^3 \right]_{\rm pert} = \frac{C_F \alpha_s}{\pi} \frac{8}{3} \mu^3 \, , 
\end{equation}
based on the consistency of \eqref{eq:perp_rel} at subleading order. 
In principle, we could now proceed and write the HQE for e.g. inclusive semileptonic $b\to c$ decays   
up to order $1/m_Q^3$. However, as we shall see below,  $\left[ \rho_D^3 \right]_{\rm pert}$ computed from the 
SV sum rules as well as from a direct calculation in HQET yield different results. Therefore, it is necessary to develop a systematic approach based on a cut-off scheme that works to all orders in $1/m_Q$. 

\subsection{Small Velocity sum rules}
Most current applications of the kinetic mass scheme make use of Small Velocity (SV) sum rules to determine the perturbative part of the HQE parameters 
\cite{Bigi:1996si,Fael:2020njb,Bigi:1994ga}. These perturbative
contributions are defined by evaluating the soft gluon contribution to a heavy-to-heavy transition, using the 
eikonal approximation for infinitely heavy quarks. The inelasticity $\omega = m_b - \sqrt{m_c^2+\vec{q}^{\, 2}} - q_0$ has for soft gluons a universal spectrum   
which, in the SV limit  $|\vec{q}| \to 0$, is given by
\begin{equation}
\frac{\text{d} I (\vec{q})}{\text{d} \omega} = I_0 (\vec{q}) \frac{4 \alpha_s}{3 \pi} C_F 
\frac{\vec{q}^{\, 2}}{2 m_c^2} \frac{1}{\omega} + \cdots \ ,
\end{equation}
where the ellipses denote higher orders in the SV expansion and $q= (q_0,\vec{q})$ is the momentum transfer to the lepton system.  
The perturbative contributions from soft gluons to the HQE parameters are related to the moments of this spectrum 
\begin{equation}
I_n (\vec{q}) = \int\limits_0^\mu \text{d}\omega  \frac{\text{d} I (\vec{q})}{\text{d} \omega}  \omega^n
= I_0 (\vec{q}) \frac{4 \alpha_s}{3 \pi} C_F \frac{\vec{q}^{\, 2}}{2 m_c^2} \int\limits_0^\mu \frac{\text{d}  \omega}{\omega} \omega^n + \cdots = I_0 (\vec{q})\frac{4 \alpha_s}{3 \pi} C_F  \frac{\vec{q}^{\, 2}}{2 m_c^2} \frac{\mu^n}{n} + \cdots\ ,
\end{equation}
expressed in terms of a cut-off $\mu$ on the inelasticity $\omega$. Note that the inelasticity is induced by the 
emission of real soft gluons, so at one loop-order a cut on $\omega$ directly corresponds to a cut $|\vec{k} | \le \mu$ on the gluon three-momentum.  
 
The general relation of the moments $I_n (\vec{q})$ to the HQE parameters are given by the SV sum rules,  \cite{Bigi:1996si,Fael:2020njb,Bigi:1994ga} 
\begin{eqnarray}
I_1 (\vec{q}) &=& I_0 (\vec{q}) \frac{\vec{q}^{\,2}}{2 m_c^2} \bar{\Lambda} + \cdots \ , \\ 
I_2 (\vec{q}) &=& I_0 (\vec{q}) \frac{\vec{q}^{\,2}}{3 m_c^2} \mu_\pi^2 + \cdots   \ ,\\
I_3 (\vec{q}) &=& I_0 (\vec{q}) \frac{\vec{q}^{\,2}}{3 m_c^2} \rho_D^3 + \cdots  \ .
\end{eqnarray}
Equating these SV sum rules to the perturbatively calculated moments within the eikonal approximation gives the perturbative contributions to the HQE parameters in this approximation. 

Working up to order $1/m^2$ this approach can be regarded as a different hard cut-off scheme, since the upper limit of the
inelasticity $\omega$ serves as the hard cut-off here. The advantage of the scheme based on the SV sum rule is that it can be 
extended to higher orders in $\alpha_s$ \cite{Fael:2020iea,Fael:2020njb}. 
However, moving on to higher orders in $1/m$ the SV sum rules 
cannot constrain all HQE parameters, since the SV moments do not contain enough information. In addition, the non-local terms appearing in the parameter $\rho^3$ in the HQET mass formula \eqref{HQETMassFomula} also cannot be accessed through an SV sum rule, so  
the cut-off scheme based on the SV sum rules cannot easily be extended to higher orders in $1/m$.

The simple hard cut-off scheme turns out to be consistent with the kinetic mass scheme based on the SV sum rules up to order $1/m^2$, but already at order $1/m^3$ inconsistencies arise between different definitions. As an example, $\rho_D^3$ can be calculated in the naive cut-off scheme in full QCD as well as in HQET at one-loop level, yielding 
\begin{eqnarray}
\left[ \rho_D^3 \right]_{\rm pert}^{SV} &=& \frac{C_F \alpha_s}{\pi}  \frac{2}{3} \mu^3\ , \\
\left[ \rho_D^3 \right]_{\rm pert}^{QCD} &=& \left[ \rho_D^3 \right]_{\rm pert}^{HQET} 
= - \frac{C_F \alpha_s}{\pi}  \frac{1}{3} \mu^3 \ ,
\end{eqnarray}
and we further note that none of these results are consistent with \eqref{eq:perp_rel}.
We conclude that an improved hard cut-off scheme is needed to clarify the situation and to open the road to go beyond $1/m^2$ for a generalized kinetic mass.

\section{Summary and Outlook} 
Increasing the precision of the HQE for inclusive semileptonic decays to the sub-percent level poses several challenges 
related to a consistent treatment of the quark mass and the HQE parameters. These problems have been identified 
already long ago, however, in practice, solving these issues remains challenging. In this paper, we focussed on the definition of the mass of the heavy quark. HQE methods expand full QCD near the mass
shell of the heavy quark. However, the on-shell quark mass is plagued by infrared renormalons leading to ambiguities expressed as a power series in 
$\Lambda_{\rm QCD} (\Lambda_{\rm QCD}/m)^n$ with $n = 0, 1,2, ...$.  To overcome this problem, a short-distance mass is defined, whose relation to the on-shell mass 
can be computed in perturbation theory. The resulting perturbative series is, however, asymptotic.

The kinetic mass is specifically designed to deal with these renormalon problems. In setting up the kinetic mass the infrared gluons from the pole mass are removed by introducing a cut-off
$\mu$ with $\Lambda_{\rm QCD} < \mu \ll m$, leaving only perturbative gluons from scales above $\mu$.  This procedure leads to a relation between the on-shell mass and the kinetic mass involving
also powers of $\mu/m$, which are related to the renormalon ambiguities. 

To access higher orders in $1/m_Q$ for the HQE it is convenient to express the HQE parameters in terms of matrix elements 
defined in full QCD, and to obtain a mass definition which is defined also in full QCD. Starting from a relation between the hadron and the quark mass derived from the forward matrix element of the trace of the energy 
momentum tensor of QCD, we obtain a differential equation for the relation between the on-shell mass and the quark mass. We use this as a definition of a generalized kinetic mass. The relation is expressed in terms of matrix elements 
in full QCD, which have to be calculated perturbatively in a cut-off scheme. In fact, the chosen scheme for the matrix 
elements further specifies the mass scheme. 

Using a naive cut-off scheme, we then obtain a relation between the extended kinetic mass and on-shell mass up to all orders in $1/m$ and to order $\alpha_s$. The result matches the calculation of the quark-self energy in the naive cut-off scheme, up to an ambiguity related to the leading renormalon, i.e. in the coefficient of the 
linear term in $\mu$.

Going beyond the leading renormalon, requires re-defining the HQE parameters accordingly. This requires a calculation of the infrared contributions to the HQE parameters, which can be done in a simple cut-off scheme to order $1/m_Q^2$. However, it turns out that the perturbative calculation of the higher order HQE parameters exhibit UV divergences. Going beyond $1/m_Q^2$, therefore, requires a new suitable hard cut-off scheme, which achieves a clean separation of IR and UV contributions. In this respect, the gradient flow seems promising. Such a scheme, in combination with our full QCD formulation of the extended kinetic mass will pave the way for a consistent treatment of sub-leading renormalons for observables of inclusive semileptonic $B$ decays.

\section*{Acknowledgements}
We thank M. Fael for useful discussions and for communicating his results on the perturbative contributions to $\rho_D^3$. We also thank P. Gambino for discussions in the early stages of this work. 
This research was supported by the Deutsche Forschungsgemeinschaft (DFG, German Research
Foundation) under grant 396021762 – TRR 257 “Particle Physics Phenomenology after the Higgs discovery''. 

This publication is part of the project “Beauty decays: the quest for extreme
precision” of the Open Competition Domain Science which is financed by the Dutch Research
Council (NWO).

\appendix

\bibliographystyle{jhep} 
\bibliography{refs.bib} 

\end{document}